\documentclass[finallayout,an,fleqn]{w-art}
\usepackage{times}
\usepackage{w-thm}
\theoremstyle{plain}

\theoremstyle{definition}

\usepackage[]{graphicx}
\chardef\bslash=`\\ % p. 424, TeXbook

\def\msun{{\,M_\odot}}

\newcommand\fsat{F_{\rm sat}}

\newcommand\sgra{Sgr~A$^*$\ }

\def\>{$>$}
\def\<{$<$}

\def\simlt{\lower.5ex\hbox{$\; \buildrel < \over \sim \;$}}
\def\simgt{\lower.5ex\hbox{$\; \buildrel > \over \sim \;$}}
\def\sqr#1#2{{\vcenter{\hrule height.#2pt
      \hbox{\vrule width.#2pt height#1pt \kern#1pt
         \vrule width.#2pt}
      \hrule height.#2pt}}}

\begin{document}
%%    The information for the title page will be placed between
%%    \begin{document} and \maketitle. The order of most entries
%%    is determined by the class file and can not be changed by
%%    rearranging them. The maketitle command follows after the
%%    abstract.
%%
%%    Most of the following commands will be completed by the publisher.
%%
%%    The copyrightyear is defined in the .clo file as the first argument
%%    of the copyrightinfo command. If the copyrightyear differs from that
%%    value it might be adjusted by the following definition:
%%
%% \renewcommand{\copyrightyear}{2002}% uncomment to change the copyrightyear.
%%
\DOIsuffix{theDOIsuffix}
%%
%% issueinfo for header and copyright line
\Volume{324}
\Issue{S1}
\Copyrightissue{S1}
\Month{01}
\Year{2003}
%%
%%    First and last pagenumber of the article. If the option
%%    'autolastpage' is set (default) the second argument may be left empty.
\pagespan{3}{}
%%
%%    Dates will be filled in by the publisher. The 'reviseddate' and
%%    'dateposted' (Published online) entry may be left empty.
\Receiveddate{15 November 2002}
\Reviseddate{30 November 2002}
\Accepteddate{2 December 2002}
\Dateposted{3 December 2002}
\keywords{}
\subjclass[pacs]{04A25}

%% \pretitle{Editor's Choice}

%% We have a short and a long form for the title. The short form
%% (optional argument) goes into the running head.

\title[Inactive Disk in Sgr~A$^*$]{The frozen (inactive) disk in
Sgr~A$^*$: freezing the accretion of the hot gas too?}

%% Please do not enter footnotes or \inst{}-notes into the optional
%% argument of the author command. The optional argument will go into
%% the header.  If there is only one address the marker \inst{x} may be
%% omitted.

%% Information for the first author.
\author[Sergei Nayakshin]{Sergei Nayakshin\footnote{e-mail: {\sf
     serg@mpa-garching.mpg.de}, Phone: +049\,089\,30000\,2258
     }\inst{1}} \address[\inst{1}]{Max Planck Institute for
     Astrophysics, Garching, Germany.}
%%
%%    Information for the second author
%\author[Sh. Second Author]{L. Second Author\footnote{Second author footnote.}\inst{1,2}}
%\address[\inst{2}]{Second address}
%%
%%    Information for the third author
%\author[Sh. Third Author]{L. Third Author\footnote{Third author footnote.}\inst{2}}
%%
%%    \dedicatory{This is a dedicatory.}
\begin{abstract}
The black hole (BH) in our Galactic Center (GC) is extremely
underluminous for the amount of hot gas available for the BH
consumption. Theoretical understanding of this fact rests on a likely
but not entirely certain assumption that the electrons in the
accreting gas are much cooler than the protons. In this case the hot
gas as a whole is too hot to accrete, and is too tenous to radiate
away its gravitational energy. Here we propose a drastically different
picture of the accretion process in Sgr~A$^*$ not based on the
unchecked two-temperature assumption. Namely, we argue that there
should exist a very cold {\em inactive} disk -- a remnant of a past
stronger accretion activity in Sgr~A$^*$. Such a disk would be a very
efficient {\em cooling surface} for the hot flow. We show that under
certain conditions the cooling due to thermal conduction cannot be
balanced by the viscous heating in the hot flow. Along with the heat,
the hot flow loses its viscosity and thus ability to accrete. It
settles (condences) onto the cold disk slighlty inside of the
circularization radius. If the latter is very large, then the
liberated energy, and the luminosity emitted, is orders of magnitude
less than naively expected. We build a simple analytical model for
this flow and calculate the expected spectra that appear to be in a
very reasonble agreement with observations. Strong additional support
for the presence of the inactive disk comes from the recent
observations of X-ray flares in Sgr~A$^*$. The properties of these
flares are very similar to those produced by stars passing through a
cold disk.
\end{abstract}
%% maketitle must follow the abstract.
\maketitle                   % Produces the title.

%% If there is not enough space inside the running head
%% for all authors including the title you may provide
%% the leftmark in one of the following three forms:

%% \renewcommand{\leftmark}
%% {First Author: A Short Title}

%% \renewcommand{\leftmark}
%% {First Author and Second Author: A Short Title}

%% \renewcommand{\leftmark}
%% {First Author et al.: A Short Title}

%% \tableofcontents  % Produces the table of contents.

\section{Introduction}

Currently, the pitiful luminosity of \sgra is most commonly explained
in the framework of Non-Radiative Accretion Flows (NRAF), a
generalization of solutions discussed in greatest detail by Narayan \&
Yi (1994; NY94 hereafter). These solutions are valid when the rate at
which the protons pass their gravitational energy to the electrons is
not significantly higher than that due to Coulomb interactions
only. This assumption is unfortunately prohibitively difficult to test
(see references in Narayan 2002). However, provided it is valid, the
electrons radiate only a tiny fraction of the total energy (NY94,
Narayan et al. 1995) in stark contrast to the standard disks (Shakura
\& Sunyaev 1973). The second important feature of NRAFs was pointed
out by Blandford \& Begelman (1999; BB99 hereafter) who showed that
these flows should produce powerful thermally driven winds (see also
Quataert, these proceedings).

In this paper we would like to point out a likely and essential
element of the accretion picture in \sgra that has so far escaped
(except for Falcke \& Melia 1997; FM97) the attention it
deserves. \sgra is believed to be closely related to the Low
Luminosity AGN (LLAGN; e.g. Ho 1999). Most if not all of these sources
seem to have cold {\em neutral and inactive} disks that often can be
seen only through water maser emission (e.g. Miyoshi et al. 1995)
arising in a range of radii where gas temperature is $200-1000$
K. Continous SED spectra of these objects also support existence of
cold disks (Quataert et al. 1999).  Because of the extremely low
ionization level of these cold disks, the disk viscosity may be nearly
zero (e.g. see Menou \& Quataert 2001). As such these ``frozen'' disks
are not {\em accretion} disks (e.g. Siemiginowska et al. 1996). Falcke
\& Melia (1997) studied the evolution of such a disk on very long time
scales in \sgra and concluded that unless the stellar wind (which is
the current source of gas for the hot flow) has a very large angular
momentum, it would have violated the near infra-red limits. In \S
\ref{sec:discussion} we argue that due to some more recent data
(e.g. Genzel 2000), the wind {\em does have} a substantial angular
momentum and hence the problem needs to be re-considered.

Consider now the implications of an inactive disk presence for the
flow of the hot gas in Sgr~A$^*$. The cool disk is razor-thin but is
much more massive than the hot flow (Nayakshin et al. 2003). For
temperatures as high as $10^7$ K, thermal conduction is usually very
important. For the conditions at hand, the time to cool off via
thermal conduction, $t_{\rm cond}$, is much shorter than that by
radiation.  Moreover, this time can even be shorter than the flow
viscous time. In this case the hot flow gets ``frozen'' by thermal
conduction before it can accrete onto the BH. As the flow looses its
thermal energy, it also looses its vertical pressure support, and
therefore it has to settle down (condensate) onto the inactive
disk. Becoming a part of the inactive disk, the gas looses its
viscosity and its ability to accrete, and can stay in the same
location essentailly indefinitely (the cold disk viscous time is as
long as $10^6 - 10^8$ years at large radii). The accretion process in
this picture would be ``delayed'' because the gas is currently pilled
up in the cold disk. The expected bolometric luminosity of this
``frozen flow'' is roughly $L_{\rm bol}\sim 0.1 \dot{M} c^2
(\hbox{few}\; R_g/R_c)$ where $R_c$ is the circularization radius of
the hot flow and $R_g= 2G M_{\rm BH}/c^2$ is gravitational radius. If
$R_c/R_g \simgt 10^{4}$, then the low luminosity of \sgra could be
understood naturally: the hot gas does not penetrate very deep into
the BH potential well, thus gaining little energy as it settles onto
the inactive disk.

%We concentrate on the simplest but likely (see \S
%\ref{sec:discussion}) scenario, that is when the angular momentum of
%the disk is aligned with that of the hot gas (i.e the both flows
%rotate in the same direction).}

During the workshop we presented 2D hydrodynamical simulations of the
hot flow above a cold inactive disk. Using an insight from our
simulations, we have recently discovered a simple idealized analytical
solution that provides a much easier understanding of the numerical
results. We present this solution below. The spectra resulting from
this condensing flow appear to agree quite well with the observational
constraints if condensation radius is $R_c\simgt 3\times 10^4 R_g$,
the disk is highly inclinded, i.e. $i\simgt 75^\circ$, and the
accretion rate in the hot flow is at its ``nominal'' value,
$\dot{M}_0\sim 3\times 10^{-6} \msun$/year.

In addition, during the meeting we realized that the observed X-ray
flares may well be due to stars passing through the inactive disk. In
Nayakshin \& Sunyaev (2003; NS03 hereafter) and in Nayakshin et
al. (2003) we calculated the expected rate of flares, duration of a
typical flare, X-ray spectra, luminosities, plus flare radio and NIR
luminosities. All of these quantities closely resemble the
observational picture reported by Baganoff et al. (2001) \& Baganoff
et al. (2003). It appears to us that both quiescent and flare spectra
of \sgra can be explained if we accept existence of an inactive frozen
disk.

\section{A simple analytical model}\label{sec:model}

A disk with temperature $T_d \simlt 100$ K and with outer radius
$R_d\simlt \;\hbox{few}\times 10^4 R_g$ could be very hard to detect
in \sgra with any current telescopes (NS03) if the disk is also highly
inclined. Since the disk is razor-thin, its viscous time scale is very
large ($\simgt 10^5$ years), and we can consider its structure as
given on shorter time scales. We thus itroduce the disk only through
the boundary conditions for the hot flow.

Two-component accretion flows are too complex to be studied
analytically except for very simplified special cases (which can
however be most insightful -- see BB99 for an example). Therefore, we
will restrict ourselves to vertically averaged equations for a
Keplerian hot flow. We consider radii $R< R_c$ where $R_c$ is the
circularization radius of the hot gas. For the low densities
concerned, the radiative cooling of the hot gas is negligible, while
the thermal conduction is at its ``best'' -- at the maximum or the
``saturated'' value. Namely, the heat flux is given by $\fsat = 5 \phi
P c_s$, where $\phi \le 1$ is the saturation parameter (Cowie \& McKee
1977), and $P$ is the vertically averaged pressure, $P=\rho c_s^2$
($\rho$ and $c_s$ are the gas density and the isothermal sound speed,
respectively). Note that if $\phi \sim 1$ then the energy flow due to
thermal conduction is effectively supersonic. This is because the flux
is carried by the electrons whose thermal speed is much higher than
that of the protons in a one-temperature plasma (for more on this see
Cowie \& McKee 1977). On the other hand, the flow of energy in the
radial direction will be proportional to the radial velocity that is
normally much smaller than the sound speed (see below). Thus the
energy losses due to thermal conduction in the limit of large $\phi$
are much larger than the rate at which the energy can be gained by
sinking in the BH potential well. That is why condensation should
ensue in this situation.

Our equations are best understood through comparison with those of
the well known ADAF solution (NY94; see their eqs. 1-4). The
stationary mass conservation equation takes into account exchange of
mass in the veritcal direction:
\begin{equation}
\frac{\partial}{\partial R}\;\left[R H \rho v_R\right]= -
\rho v_z R\;,
\label{massc}
\end{equation}
where $v_R$ is the radial flow velocity ($v_R>0$ for accretion), $\rho
v_z$ is the mass flow density for condensation ($v_z<0$) or
evaporation ($v_z>0$). The vertical scale height $H$ is introduced
through the hydrostatic balance as $H=c_s/\Omega$ where $\Omega$ is
the angular velocity. Since the latter is assumed to be Keplerian, the
radial momentum equation (eq. 2 in NY94) is trivially satisfied.
Since the cold disk is also Keplerian, there is no exchange of
specific angular momentum between the two flows and the angular
momentum conservation equation (3 in NY94) is unaltered. With
$\Omega=\Omega_K$ we get
\begin{equation}
\dot{M} = 4\pi R H \rho v_R = \frac{12\pi \alpha}{R \Omega_K} 
\frac{\partial}{\partial R}\;\left[  \rho c_s^2 R^2 H \right]\;,
\label{vr}
\end{equation}
where $\alpha$ is viscosity parameter (Shakura \& Sunyaev 1973).  The
entropy equation should include (in addition to the usual terms) the
thermal conduction flux, $\fsat$, and the hydrodynamical flux of
energy in the vertical direction.  To derive this equation, we follow
the formalism of Meyer \& Meyer-Hofmeister (1994; see their equation
8), with the following exceptions. We neglect winds here because we
are interested in cooler, condensing solutions. Thus their side-way
term (their eq. 5) is not included. In addition, the radial entropy
flow term (the ``advective cooling''; NY94) is designated $Q_{\rm
adv}$. Following NY94 we set $Q_{\rm adv} = f_{\rm adv} Q_+$, where
$f_{\rm adv} \le 1$ is a parameter. Further, for subsonic flows
$v_z^2\ll c_s^2$, and we obtain
\begin{equation}
Q_+ - \fsat - \rho v_z \left[ \;\frac{5}{2}c_s^2\; + \;\frac{GM_{\rm
BH}}{2R}\; \left(\frac{H}{R}\right)^2 \right] - Q_{\rm adv} = 0\;,
\label{eflux}
\end{equation}
where $Q_+ = (9/2) \alpha \rho c_s^3$ is the viscous heating rate.
The factor of $5/2$ in eq. (\ref{eflux}) is $\gamma/(\gamma-1)$ for
the $\gamma=5/3$ gas that we consider here.

After some simple algebra and using $GM_{\rm BH}H^2/2R^3 = c_s^2/2$,
we find that the condensation velocity is
\begin{equation}
v_z = - \left(\frac{5}{3}\phi - \frac{3}{2}\alpha (1-f_{\rm
adv})\right) c_s = - b c_s\;,
\label{vz}
\end{equation}
where $b$ is introduced for convenience. The equation (\ref{vz}) is
crucial for the rest of the paper. If thermal conduction is vigorous,
i.e. $b>0$, then the hot flow is condensing onto the cold disk. In the
opposite case of a small $\phi$ and a ``large'' $\alpha$, $b<0$,
viscous heating prevails. Thermal conduction then serves to evaporate
the inactive disk. We should also note that the regime of large
$\alpha$ ($\sim 0.3$) was already studied by F. Meyer and
collaborators in many papers. Their solutions are for higher
accretion rates and therefore they are in the non-saturated regime,
which is roughly speaking equivalent to the $\phi\ll 1$ case. As they
found, the {\rm evaporation} is a very strong function of $\alpha$
($\propto \alpha^3$). Our results (eq. \ref{vz} in particular) are
thus in a complete agreement with that of Meyer \& colleagues, and we
essentially extend their work on the case $\phi \gg \alpha$.

Inserting now $v_z=-bc_s$ into equation (\ref{massc}) and also
substituting $v_R$ on its value found from equation (\ref{vr}), we
arrive at a second order differential equation that contains two
variables, $\rho$ and $c_s$:
\begin{equation}
b \rho R c_s = 3 \alpha \frac{\partial}{\partial R}\;\left\{\;
\frac{1}{R \Omega_K}\; \frac{\partial}{\partial R}\; \left[\rho c_s^3
\frac{R^2}{\Omega_K}\right]\right\}\;.
\label{long}
\end{equation}
This equation cannot be solved in a general case. By introducing
$v_z\ne 0$ we added an extra variable to the accretion flow equations,
and the number of independent equations is now smaller than the number
of unknowns. (In particular, both $v_z$ and $c_s$ are to be found from
the single energy equation \ref{eflux}). This situation is well known
in analytical ADIOS wind solutions (BB99). In
the latter case one has to introduce three free parameters that
describe the mass, energy and angular momentum carried away by the
wind.

The most natural way to proceed here is to suggest that the
temperature is a power-law function of radius. For example, for ADAF,
$T(R)\propto R^{-1}$ in a broad range of radii. On the other hand,
thermal conduction tends to smooth out temperature gradients (for
example within supernova remnants), and hence in the other extreme
$T(R)\simeq $~const (we in fact observed this nearly constant $T(R)$
in our numerical simulations). Thus, $c_s = c_{0} (R_c/R)^{\lambda}$,
where $0\le \lambda\le 1/2$ and $c_{0}$ is the sound speed at
$R_c$. If we define $u\equiv \sqrt{R/R_c}$, then equation (\ref{long})
can be re-written as
\begin{equation}
\rho u^{3 -2 \lambda} = \frac{3 \alpha}{4b}\; \frac{R_c c_0^2 }{G
M_{\rm BH}}\; \frac{\partial^2 }{\partial u^2}\;\left[\rho
u^{7-6\lambda}\right] \;.
\label{notlong}
\end{equation}
Finally, defining $\tilde{\rho}\equiv \rho u^{7-6\lambda}$, we obtain
\begin{equation}
\frac{\partial^2 \tilde{\rho}}{\partial u^2} = \frac{1}{l_c^2}\;
\frac{\tilde{\rho}}{u^{4(1-\lambda)}}\;.
\label{simple}
\end{equation}
Here $l_c$ is the dimensionless ``condensation length'':
\begin{equation}
l_c^2 \equiv \frac{3 \alpha}{4b}\;\frac{ c_{0}^2 R_c}{G M_{\rm BH}}\;.
\label{lc}
\end{equation}
If there were no thermal conduction losses, the internal energy of the
hot gas would be of order its gravitational energy. If the inactive
disk extends to $R > R_c$, the thermal conduction will reduce the gas
thermal energy. Hence we expect that $c_{0}^2 R_c/G M_{\rm BH}\le 1$.

We are most interested in the case $\phi \gg \alpha$ and therefore we
shall only explore the $\lambda=0$ case below.\footnote{Nevertheless
we note that (i) approximate solutions may be obtained for a general
value of $\lambda$, and (ii) for $\lambda=1/2$ there is a scale-free
solution.} In addition, with $\alpha\ll \phi$, condensation length is
small, i.e. $l_c^2 < \alpha/b \sim \alpha/\phi \ll 1$. This
circumstance facilitates finding an approximate analytical solution for
equation (\ref{notlong}):
\begin{equation}
\tilde{\rho} = \;\hbox{const}\; \exp\left[-\frac{1}{l_c u}\right]\;.
\label{approx1}
\end{equation}
Indeed, $d^2\tilde{\rho}/du^2 = \tilde{\rho}\;(-2/l_cu^3 +
1/l_c^2u^4)\simeq \tilde{\rho}/l_c^2u^4$ due to the fact that $1/l_c u
\gg 1$. Let us now explicitly write down the results from this
approximate solution:
\begin{eqnarray}
\rho(R) = \rho_c \left(\frac{R_c}{R}\right)^{7/2}\;
\exp\left[-\frac{1}{l_c}\left(\sqrt{\frac{R_c}{R}}-1\right)\right]\;,
\label{densr}\\
v_R \;=\; \frac{3 \alpha}{2 l_c}\; \frac{c_0^2}{R_c \Omega_K(R_c)}
\;=\; \sqrt{\frac{\alpha b}{3}}\;c_0 \; =\; \hbox{const}\ll c_0\;,
\label{vrr}\\
\dot{M}(R)\;=\; \dot{M}_0 \frac{R_c}{R}
\exp\left[-\frac{1}{l_c}\left(\sqrt{\frac{R_c}{R}}-1\right)\right]\;.
\label{mdotr}
\end{eqnarray}
Here $\rho_c$ is gas density at $R_c$, and $\dot{M}_0$ is the
accretion rate at that point. Note that equation (\ref{vrr}) shows
that the radial velocity is constant and is substantially smaller than
the sound speed. Further, for $l_c< 1/2$ (recall that we assumed
$l_c\ll 1$), the accretion rate increases with $R$, as it should for a
condensing flow.

\section{Sample results}\label{sec:results}

\begin{figure}[htb]
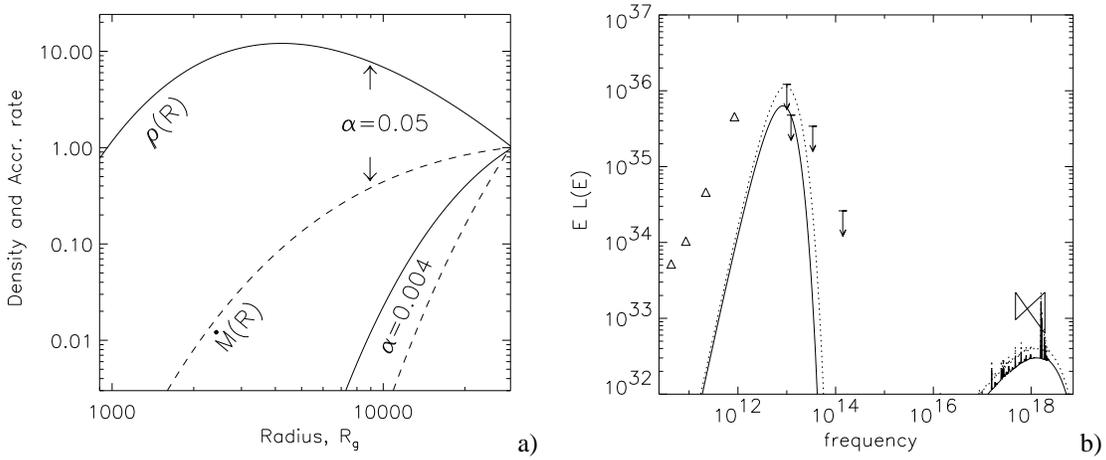

\includegraphics[width=.45\textwidth]{paper1a.epsi}~a)
\hfil
\includegraphics[width=.45\textwidth]{paper1b.epsi}~b)
\caption{(a) -- Radial density (solid) and accretion rate (dashed)
profiles for $\phi=0.2$, $R_c=3\times 10^4 R_g$, and two values of
viscosity parameter $\alpha = 0.05$ (upper curves) and 0.004 (lower
curves). (b) -- Spectra corresponding to the radial profiles shown in
(a). The dotted and solid curves are for $\alpha=0.05$ and
$\alpha=0.004$, respectively. For both cases inclination angle
$i=75^\circ$ and accretion rate $\dot{M}_0=3\times 10^{-6}\msun$
year$^{-1}$.  The time-averaged spectrum of \sgra is shown with
triangles (radio data; detections), upper limits (infra-red) and the
bow-tie (X-rays; Chandra data). The lower frequency data are from
references given in Melia \& Falcke (2001), except for the 2.2 $\mu$m
point that is from Hornstein et al. (2003), and for Chandra data that
are from Baganoff et al. (2003). For the X-ray part of the data the
spectrum should only be lower than the Chandra observations because
the latter are probably dominated by the emission of hot gas at $R\sim
R_B > R_c$, which is not included in the model.}
\label{fig:1}
\end{figure}

To illustrate the analytical solution presented above, we plot the gas
density and the mass accretion rate profiles (equations \ref{densr} \&
\ref{mdotr}) in Figure \ref{fig:1} for $\phi=0.2$ and two different
values of $\alpha$. These are chosen to represent the two opposite
extremes. In the case of larger $\alpha$, condensation length is
$l_c=0.38$ and this is barely satisfies the condition $l_c\ll 1$ under
which our approximate solution is valid. The accretion rate changes
slowly with radius in this case, meaning that condensation is
relatively slow. In the case of very small $\alpha$ the hot flow
collapses onto the disk far quicker: the mass flow is reduced by $\sim
100$ times already at $R\simeq R_c/2$.  This is easily understood by
noting that $v_z/v_R = \sqrt{3b/\alpha} \simeq 4$ for the larger
$\alpha$ value, whereas for $\alpha=0.004$ this ratio is $v_z/v_R
\simeq 16$. The large value of the ratio $v_z/v_R$ in our solutions
justifies the statement that we made in the introduction:
``condensation time'', $t_{\rm cond} = H/v_z$, is much shorter than
the viscous time, $t_{\rm visc} = R/v_R$ for our solutions.

\begin{figure}[htb]
%\vskip0.3cm
\begin{minipage}{.45\textwidth}
\includegraphics[width=\textwidth]{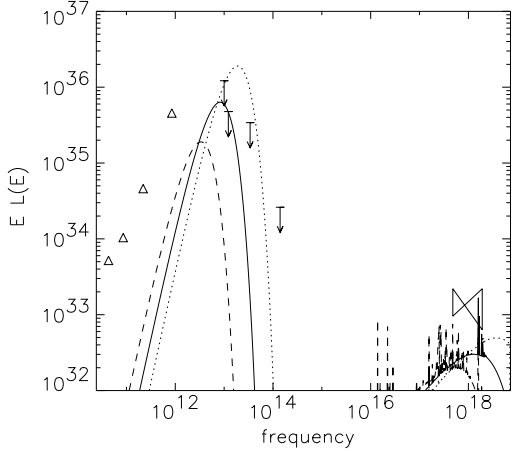}
\label{fig:2}
\end{minipage}
\hfil
\begin{minipage}{.45\textwidth}
\setfloattype{figure}
\caption{Spectrum emitted by the condensing flow for inclination angle
$i=75^\circ$, accretion rate $\dot{M}_0=3\times 10^{-6}\msun$ and
three values of circularization radius $R_c$: $10^4 R_g$ (dotted),
$3\times 10^4 R_g$ (solid), and $10^5 R_g$ (dashed), respectively. As
expected, the bolometric luminosity increases as $R_c$ decreases
because the hot gas gains more gravitational energy for larger $R_c$.}
\end{minipage}
\end{figure}

It is useful to define the ``apparent'' bolometric efficiency
coefficient of the frozen accretion flows, $\varepsilon_{\rm bol}$ in
the standard way:
\begin{equation}
\varepsilon_{\rm bol} \equiv \frac{L_{\rm bol}}{\dot{M}_0 c^2}\;,
\label{edef}
\end{equation}
where $L_{\rm bol}$ is the bolometric luminosity of the source. One
should recall that for standard accretion flows, $\varepsilon_{\rm
bol}\sim 0.1$ for a non-rotating BH.  The energy deposition into the
cold disk is given by the second and the third terms in equation
(\ref{eflux}). Integrating this expression over the disk surface area
from $3 R_g$ to $R_c$, one can arrive at $\varepsilon_{\rm bol} = 1.5
(1 + \zeta) [1 + 2 l_c + 2 l_c^2] (R_g/R_c)$, where $\zeta\equiv
(1-0.9 \alpha/\phi)^{-1}$. For $\phi\gg \alpha$ we have $\zeta \simeq
1$ and $l_c\ll 1$, so $\varepsilon_{\rm bol}\simeq 3 R_g/R_c$. This
expression is of course only applies at $R_c \gg 3 R_g$ because we
neglected any relativistic corrections to the gravitational
potential. In addition, the derived expression is only a rough
estimate since we assumed vertically averaged equations which are
clearly inaccurate for $H\sim R$ (which in our simple model occurs at
$R = R_c$). One can in fact show that realistically $\varepsilon_{\rm
bol}$ should be smaller by a factor of at least few. The point here is
that our simplistic solution is not bound at $R=R_c$ because its
thermal energy exceeds gravitational energy at that point. So either a
wind takes away the excess energy (as BB99 argued for ADAF solutions)
or more likely this simply means that we over-estimated the gas
temperature at $R_c$. We should also explicitly insert the disk
inclination angle into the definition of the apparent efficiency since
the observed luminosity of the thin disk is approximately proportional
to $\cos i$. Variation of the predicted spectrum with $R_c$ is shown
in Figure (\ref{fig:2}). Summarizing, the apparent bolometric
efficiency of the freezing flow is
\begin{equation}
\varepsilon_{\rm bol} \simlt \; \cos{i} \;\frac{R_g}{R_c} \; = 10^{-4}\;
\cos{i} \;\frac{10^4 R_g}{R_c}
\label{eval}
\end{equation}
In a similar spirit, the X-ray radiative efficiency can be
defined. Approximately, $\varepsilon_{\rm x} \sim (R_g/R_c)\; (t_{\rm
cond}/t_{\rm cool})$, where $t_{\rm cool}$ is the radiative cooling
time for the hot flow. For Sgr A$^*$, ratio $t_{\rm cond}/t_{\rm
cool}$ is extremely small and this is why X-ray contribution to the
radiative output of \sgra is so small.

%\begin{vchfigure}%[htb]
%  \includegraphics[width=.45\textwidth]{paper2b.epsi}
%\vchcaption{Spectrum emitted by the condensing flow for inclination
%angle $i=75^\circ$, accretion rate $\dot{M}_0=3\times 10^{-6}\msun$
%and three values of circularization radius $R_c$: $10^4 R_g$ (dotted),
%$3\times 10^4 R_g$ (solid), and $10^5 R_g$ (dashed), respectively.}
%\label{fig:2}
%\end{vchfigure}

\section{Discussion}\label{sec:discussion}

We have suggested here that there exist an inactive disk around
Sgr~A$^*$, a remnant of a past powerful accretion (and probably star
forming) activity. The disk may be quite light compared with both the
BH and the star cluster (Nayakshin et al. 2003), yet it easily
out-weights the $\sim 10^{-3}\msun$ of the hot gas present in the
region interior to the Bondi radius. The disk then serves as a very
efficient cooling surface for the hot flow. The flow essentially gets
{\em frozen} (stopped), and its energy is radiated as thermal emission
at frequencies much below the X-ray band. The X-ray emitting flow thus
simply disappears from ``the radar screen''. This in our opinion may
be the explanation of the exceptional {\em apparent} X-ray radiative
inefficiency of Sgr A$^*$. Further, since the hot flow does not
penetrate very deep into the BH potential well, its total bolometric
luminosity is $\sim \;\hbox{few}\; R_g/R_c$ times smaller than that
expected if the gas made it all the way into the BH and was
radiatively efficient. Thus the flow also {\em appears} radiatively
inefficient in the bolometric sense. As we explained in the paper, we
believe that the accretion of the winds from the hot stars is simply
delayed in time and it is by no means radiatively inefficient in the
long run.

Falcke \& Melia (1997; FM97) have assumed the hot wind infall as given
and studied viscous evolution of the ``fossil'' disk on long time
scales, whereas we concentrate on much shorter time scales on which
the structure of the disk does not change (i.e $t < t_{\rm visc}\sim
10^6 \alpha^{-1}$ years for $T_{\rm disk}=100$ K and $R=10^4 R_g$).
Our study is therefore complimentary to that of FM97.

Our results concerning the conditions under which the wind--disk (or
the hot flow--disk) interactions will not violate the tight NIR limits
are quite similar to that of FM97. In particular, FM97 note that ``the
Bondi-Hoyle wind must be accreting with a very high specific angular
momentum to prevent it from circularizing in the inner disk region
where its impact would be most noticeable''. We find that the
circularization radius should be $\simgt 3\times 10^4 R_g$, implying a
very large angular momentum indeed. Further, we suggested that the
disk and the hot flow angular momenta are at least approximately
aligned or else there would be a substantial heating due to friction
between the two, a heating not included in our analysis. It remains to
be seen whether results will be qualitatively similar if the disk and
the hot flow rotation axises are misaligned.

FM97 considered a ``large'' value of $R_c$ being rather unlikely. We
however note that according to recent data, the stars from which the
hot wind originates appear to be on tangential orbits counter-rotating
the Galactic rotation and are $\sim \;\hbox{few arcsecond}\; \sim
\hbox{few}\times 10^5 R_g$ off \sgra (e.g. Genzel 2000). There is thus
no deficit of angular momentum at these distances. Finally, we only
studied here the region of the flow interior to $R_c$. However the
exchange of the angular momentum between the hot flow and the disk
should take place at $R> R_c$ where the hot gas is sub-Keplerian. This
should enrich the hot flow with the angular momentum directed as that
of the the disk and hence the circularization of the hot flow should
occur even if it had zero angular momentum at infinity. Thus the
requirement of a large angular momentum in the wind may be relaxed,
although this effect remains to be quantified with future
calculations.

\section{Conclusion}

In this paper we suggested that there exists a very cold inactive disk
in Sgr~A$^*$, and that its role in the accretion picture is
significant. While the hot gas is very tenuous and cannot radiate its
energy away, it can easily transfer its energy into the cold disk via
thermal conduction. The cold disk is much denser and much more massive
than the hot flow and can serve as a very powerful freezer (or
radiator) for the hot flow. As the hot flow looses its energy, it also
looses its viscosity and ``sticks'' to the cold disk.  The accretion
flow is thus quenched by this seemingly {\em ``non-radiative''}
cooling. One can easily check that neither internal viscous
dissipation nor the heat input from the hot flow in \sgra are
sufficient to overcome the radiative cooling and restart the accretion
in the inactive disk. It appears that only arrival of a new large
supply of low angular momentum material could revive the inactive disk
in \sgra now.

We thank H. Falcke, C. McKee, F. Meyer, R. Narayan, R. Sunyaev and
H. Spruit for discussions.% that aided this research.


\begin{thebibliography}{10}

\bibitem[]{} Baganoff F.~K., et~al., 2001, Nature,
413, 45

\bibitem[]{} Baganoff F.~K., et~al. 2003, these proceedings

\bibitem[]{} Blandford, R., \& Begelman, M.C., 1999, MNRAS, 303, L1

\bibitem[]{} Cowie, L.L., \& McKee, C.F. 1977, ApJ, 211, 135

\bibitem[]{} Falcke, H, \& Melia, F. 1997, ApJ, 479, 740

\bibitem[]{} Genzel, R. 2000, in the Proceedings of the Star2000
Meeting, editor R. Spurzem (astro-ph/0008119)

\bibitem[]{} Hornstein, S.D. et al. 2003, these proceedings.

%\bibitem[]{} Liu, S., \& Melia, F. 2002, ApJ, 566, L77

%\bibitem[]{} {Markoff} S., et al. 2001, A\&A, 379, L13

\bibitem[]{} Melia, F., \& Falcke, H. 2001, ARA\&A, 39, 309

\bibitem[]{} Menou, K., \& Quataert, E. 2001, ApJ, 552, 204

\bibitem[]{} Meyer, F., \& Meyer-Hofmeister, E. 1994, A\&A, 288, 175

\bibitem[]{} Miyoshi, M., et al. 1995, Nature, 373, 127

\bibitem[]{} Narayan, R., \& Yi, I. 1994, ApJ, 428, L13

\bibitem[]{} Narayan, R., Yi, I., \& Mahadevan, R. 1995, Nature, 374,
623

\bibitem[]{} {Narayan} R., 2002, p. 405, in ``Lighthouses of the
Universe'', Springer 2002, editors Gilfanov, M., Sunyaev, R., \&
Churazov E.

%: The Most Luminous
%Celestial Objects and Their Use for Cosmology 


\bibitem[]{} Nayakshin, S., \& Sunyaev, R. 2003, submitted to MNRAS
(astro-ph/0302084)

\bibitem[]{} Nayakshin, S., et al. 2003, in preparation.

\bibitem[]{} Quataert, E., et al. 1999, ApJL, 525, L89

\bibitem[]{} {Sch\"odel} R., et al. 2002, Nature, 419, 694

\bibitem[]{} Shakura, N.I., \& Sunyaev, R.A. 1973, A\&A, 24, 337

\bibitem[]{} Siemiginowska, A., Czerny, B., \& Kostyunin, V. 1996,
ApJ, 458, 491

\end{thebibliography}
\end{document}